\documentclass[onecolumn]{emulateapj}

\usepackage{natbib,amsmath}

\newcommand{\msol}{M_\odot}

\newcommand{\kms}{km~s$^{-1}$}
\newcommand{\lya}{Ly$\alpha$}

\def\h2{H$_2$}
\def\f0{$F_0$}

\newcommand{\cm}[1]{\, {\rm cm^{#1}}}
\newcommand{\mkms}{{\rm \; km\;s^{-1}}}

\newcommand{\nion}{N$^{+4}$}
\newcommand{\snt}{\sigma({\rm N^{+3}})}
\newcommand{\nthr}{N$^{+3}$}
\newcommand{\N}[1]{{N({\rm #1})}}
\newcommand{\sci}[1]{{\rm \; \times \; 10^{#1}}}

\newcommand{\nhi}{$N_{\rm HI}$}

\def\ltk{\left [ \,}
\def\ltp{\left ( \,}

\def\rtk{\, \right  ] }
\def\rtp{\, \right  ) }

\def\nhi{$N_{\rm HI}$}

\begin{document}

\title{A Survey for NV Absorption at $z \approx z_{GRB}$
in GRB Afterglow Spectra:
Clues to Gas Near the Progenitor Star}
%\title{NV Gas in GRB Afterglow Spectra:
%A Snapshot of Gas Near the Progenitor}
%\author{Blah}
\author{Jason X. Prochaska\altaffilmark{1,2},
	Miroslava Dessauges-Zavadsky\altaffilmark{3},
        Enrico Ramirez-Ruiz\altaffilmark{2}, 
        Hsiao-Wen Chen\altaffilmark{4}
}
\altaffiltext{1}{University of California Observatories - 
Lick Observatory, University of California, Santa Cruz, CA 95064} 
\altaffiltext{2}{Department of Astronomy and Astrophysics,
University of California, Santa Cruz, Santa Cruz, CA 95064} 
\altaffiltext{3}{Observatoire de Gen\`eve, 51 Ch. des Maillettes, 
1290 Sauverny, Switzerland}
\altaffiltext{4}{Dept.\ of Astronomy \& Astrophysics and
Kavli Institute for Cosmological Physics,
5640 S.\ Ellis Ave, Chicago, IL, 60637, U.S.A.}

\begin{abstract}
We survey \ion{N}{5} absorption in the afterglow spectra of 
long-duration gamma-ray bursts (GRBs) with the intent to study highly
ionized gas in the galaxies hosting these events.
We identify a high incidence (6/7) of spectra exhibiting
\ion{N}{5} gas with $z \approx z_{GRB}$
and the majority show large column densities
$\N{N^{+4}} \gtrsim 10^{14}\cm{-2}$.  
With one exception, 
the observed line-profiles
are kinematically `cold', i.e.\ they are
narrow and have small velocity offset
($\delta v \lesssim 20 \mkms$) from absorption lines associated
with neutral gas.  
In addition, the \ion{N}{5} absorption has similar velocity as
the UV-pumped fine-structure lines indicating these high ions are
located within $\approx 1$\,kpc of the GRB afterglow.
These characteristics are unlike those for 
\ion{N}{5} gas detected in the
halo/disk of the Milky Way or along sightlines through 
high $z$ damped \lya\ systems
but resemble the narrow absorption line systems associated 
with quasars and some high $z$ starbursts. 
We demonstrate that 
GRB afterglows photoionize nitrogen to \nion\ at
$r \approx 10$\,pc. 
This process can produce \ion{N}{5} absorption with characteristics
resembling the majority of our sample and 
and we argue it is the
principal mechanism for \nion\ along GRB sightlines.
Therefore, the observations provide a snapshot of the 
physical conditions at this distance. 
In this scenario, the observations
imply the progenitor's stellar wind is confined to $r < 10$\,pc
which suggests the GRB progenitors occur within dense ($n > 10^3 \cm{-3}$)
environments, typical of molecular clouds.  
The observations, therefore, primarily constrain the 
physical conditions -- metallicity, density, velocity fields -- of
the gas within the (former)
molecular cloud region surrounding the GRB.  

\end{abstract} 

\section{Introduction}

Long-duration gamma-ray bursts (GRBs) are believed to have massive
star progenitors arising in active star-forming regions of high $z$
galaxies \citep[e.g.][]{wb06}.  Roughly half of these events have
associated UV/optical afterglows and a subset of these have apparent
magnitudes sufficient for high-resolution spectroscopy using 10m-class
telescopes \citep[e.g][]{fdl+05,cpb+05}.  In principle, the power-law
afterglow spectrum has imprinted within it features from gas
throughout the interstellar medium (ISM) of the host galaxy.
%surrounding the progenitor, its associated
%star-forming region, and the ambient interstellar medium (ISM) 
%of the host galaxy.
This stands in contrast to studies of quasars
whose integrated photon output ionizes their ISM\footnote{The obvious 
exception is the gas identified as broad absorption
lines in quasar spectra which is located very close to the quasar.}
and surrounding gas out to many tens of kpc. % \citep[e.g.][]{cmb+08}.
Furthermore, although quasar sightlines frequently penetrate
foreground, star-forming galaxies 
\citep[the so-called damped \lya\ systems, QSO-DLA;][]{wgp05},
these are probed according to gas cross-section and quasar sightlines
should only rarely intersect the small, dense
regions undergoing active star-formation \citep{zp06}.

In these respects, GRB afterglow spectra allow one to probe a
diversity of phases in the ISM of star-forming galaxies: the
circumstellar material from the massive star progenitor, the
\ion{H}{2} region produced by the progenitor and neighboring OB stars,
the neutral ISM of the host galaxy, and any diffuse gas within the
galactic halo.  Unfortunately, even though these phases arise at
distinct distances along the sightline, the observed spectrum resolves
only the relative velocities of the gas.  To focus analysis on a
specific phase, one is generally forced to isolate a unique ion and/or
material associated with a specific velocity.

GRB afterglow spectra reveal large column densities of \ion{H}{1} gas and metals
associated with the host galaxy ISM \citep{sff03,vel+04,jfl+06}.  The
analysis of the metal-line transitions have localized this neutral gas
within the ambient ISM of the host galaxy.  Specifically, the
detection of fine-structure lines of Si$^+$ and Fe$^+$ ions places the
gas within $\approx 1$\,kpc of the GRB afterglow while the detection
of \ion{Mg}{1} absorption requires the gas to lie at distances greater
than $\approx 100$\,pc \citep{pcb06}.  These conclusions are supported
by direct distance determinations based on analysis of
line-variability in fine-structure lines \citep[$\approx 100$\,pc and
  2\,kpc from GRB~020813 and GRB~060418
  respectively;][]{dcp+06,vls+07}.  The majority of afterglow
spectra also show
high-ion absorption (e.g.\ \ion{C}{4}) that is offset by several tens
to hundred \kms\ from the peak optical depth of the neutral ISM
\citep{cpr+07}.  This diffuse, ionized gas is also traced by strong
transitions of low-ions (e.g.\ \ion{Si}{2}~1526) but without
coincident fine-structure absorption.  Therefore, the gas must lie at
distances greater than a few kpc from the GRB.  These characteristics
identify the clouds as partially ionized gas within the halo of the
GRB host galaxy \citep{pcw+08}.

While studies of the gas in the neutral ISM and galactic halo are
valuable for studying the physical conditions in star-forming
galaxies, these phases offer only indirect constraints on the nature
of the GRB progenitor \citep{rtb02}. 
Of great interest is to identify gas located within the star-forming
region or even gas shed by the progenitor itself. To date, however, no
study has presented compelling evidence for gas within $\approx
100$\,pc of the GRB: neither circumstellar material \citep{cpr+07},
the molecular cloud that presumably beget the progenitor
\citep{tpc+07}, nor material associated with a pre-existing \ion{H}{2}
region \citep{wph+08}.  Regarding the latter phase, most of the key diagnostics
(e.g.\ \ion{Si}{4}, \ion{Al}{3}, \ion{Si}{3}, \ion{C}{4}) are either
compromised by blending with the \lya\ forest or can be confused with
a galactic halo component.  Presently, there is only indirect evidence
for significant column densities of ionized gas near GRB: a number of
GRB sightlines exhibit X-ray absorption with implied metal column
densities that significantly exceed the neutral ISM column densities
measured from the optical spectra \citep{gw01,whf+07}.  
This result\footnote{There have also been claims of 
temporal variation in the X-ray absorption spectrum
\citep[e.g.][]{clr+07} which would also imply significant
metals near the GRB afterglow \citep{lp02}, but these variations
are more naturally explained by the temporal evolution of the
intrinsic X-ray spectrum \citep{bk07}.}
hints at a large reservoir of highly ionized gas near the GRB which
has not yet been revealed by the rest-frame UV spectra acquired by
ground-based facilities.

The challenge to identify and study gas close to the GRB has motivated
us to survey GRB afterglow spectra for the presence of \ion{N}{5}
absorption.  Because the \nthr\ ion has a large ionization potential
(IP=77eV), it is difficult to produce \nion, especially using stellar
radiation fields.  In the ISM of local galaxies, \nion\ is generally
believed to trace collisionally ionized gas either in equilibrium at a
high temperature ($T > 10^{5}$K) or out of equilibrium due to a
post-shocked gas cooling from $T > 10^6$\,K \citep[e.g.][]{is04a}.  In
terms of GRB studies, however, the GRB event itself and its bright
afterglow emit sufficient numbers of $h\nu \approx 80$\,eV photons to
produce N$^{+4}$ gas near the progenitor.  Observationally, the
\nion\ ion is notable for exhibiting an alkali doublet at
$\lambda\lambda 1238,1242$ in the rest-frame which lies redward of the
\ion{H}{1} \lya\ transition.  In GRB sightlines, therefore, this
transition (unlike \ion{O}{6} and \ion{S}{6} doublets)
does not blend with \ion{H}{1} absorption from the
\lya\ forest, and GRB afterglow spectra that cover \lya\ will often
provide an analysis of the \ion{N}{5} doublet.  Indeed, previous
studies have reported the detection of the \ion{N}{5} doublet along
individual sightlines \citep{vel+04,cpb+05,twl+07}.

In this paper we perform a systematic search and analysis
of \nion\ gas for a modest sample of $z>2$ GRBs.  We 
report on the incidence of its detection, its characteristic
column density, and compare the line-profiles with 
other transitions identified along the sightline.  
Finally, we investigate
the origin of this gas and explore constraints on
the nature of the GRB progenitor environment.

\begin{deluxetable}{lcccccc}
\tablewidth{0pc}
\tablecaption{GRB-NV SAMPLE\label{tab:obs}}
\tabletypesize{\footnotesize}
\tablehead{\colhead{GRB} &\colhead{RA} & \colhead{DEC} & \colhead{Instrument} & 
\colhead{$R$} & \colhead{Ref}}
\startdata
\object{021004}&00:26:54.68&+18:55:41.6&UVES&52,000&1\\
\object{030323}&11:06:09.40&$-$21:46:13.2&FORS2&2,600&2\\
\object{050730}&14:08:17.14&$-$03:46:17.8&MIKE&30,000&3\\
\object{050820}&22:29:38.11&+19:33:37.1&HIRES&30,000&4\\
\object{050922C}&19:55:54.48&$-$08:45:27.5&UVES&30,000&5\\
\object{060206}&13:31:43.42&+35:03:03.6&ISIS&4,000&6\\
\object{060607}&21:58:50.40&$-$22:29:46.7&UVES&43,000&7\\
\enddata
\tablerefs{
1: \cite{fdl+05};
2: \cite{vel+04};
3: \cite{cpb+05};
4: \cite{pcb+07};
5: \cite{pwf+07};
6: \cite{twl+07};
7: \cite{gcn5237}}
 
\end{deluxetable}

\begin{deluxetable*}{lcccccccccccc}
\tablewidth{0pc}
\tablecaption{SURVEY SUMMARY\label{tab:summ}}
\tabletypesize{\footnotesize}
\tablehead{\colhead{GRB} &\colhead{$z_{GRB}$} & \colhead{log \nhi} & 
\colhead{[M/H]$^a$} & \colhead{$W_{1238}^b$} & 
\colhead{$\log \N{N^{+4}}$} & \colhead{$\delta$(NV)$^c$} \\
& & ($\cm{-2}$) &  & (\AA) & ($\cm{-2}$) & (\kms) }
\startdata
GRB021004&2.3291&19.00&$ 0.0$&$ 0.307 \pm 0.008$&$ 14.64 \pm 0.04$&$-10\pm 10$\\
GRB030323&3.3720&21.90&$>-0.9$&$ 0.325 \pm 0.043$&$> 14.35$&$ 20\pm 30$\\
GRB050730&3.9686&22.15&$-2.3$&$ 0.142 \pm 0.021$&$ 14.09 \pm 0.08$&$  5\pm  5$\\
GRB050820&2.6147&21.00&$-0.6$&$ 0.045 \pm 0.007$&$ 13.45 \pm 0.05$&$-90\pm 10$\\
GRB050922C&2.1990&21.60&$-2.0$&$ 0.197 \pm 0.026$&$> 14.19$&$-20\pm  5$\\
GRB060206&4.0480&20.85&$-0.9$&$ 0.093 \pm 0.010$&$ 13.73 \pm 0.15$&$  0\pm 15$\\
GRB060607&3.0748&16.80&$ 0.0$&$-0.009 \pm 0.003$&$< 12.61$&$$\\
\enddata
\tablenotetext{a}{Gas metallicity derived from low-ion absorption. See \cite{pcd+07} and \cite{dpc+08} for details.}
\tablenotetext{b}{Rest-frame equivalent width of the \ion{N}{5}~1238 transition.}
\tablenotetext{c}{Estimated velocity offset between the rough centroid of the \ion{N}{5} line-profile and the peak optical depth of the fine-structure lines.}
 
\end{deluxetable*}

\begin{figure*}
\epsscale{0.8}
%\plotone{../Figures/fig_data.ps}
%\centerline{\epsfxsize=\hsize{\epsfbox{f1.eps}}}
\begin{center}
\includegraphics[height=6.8in,angle=90]{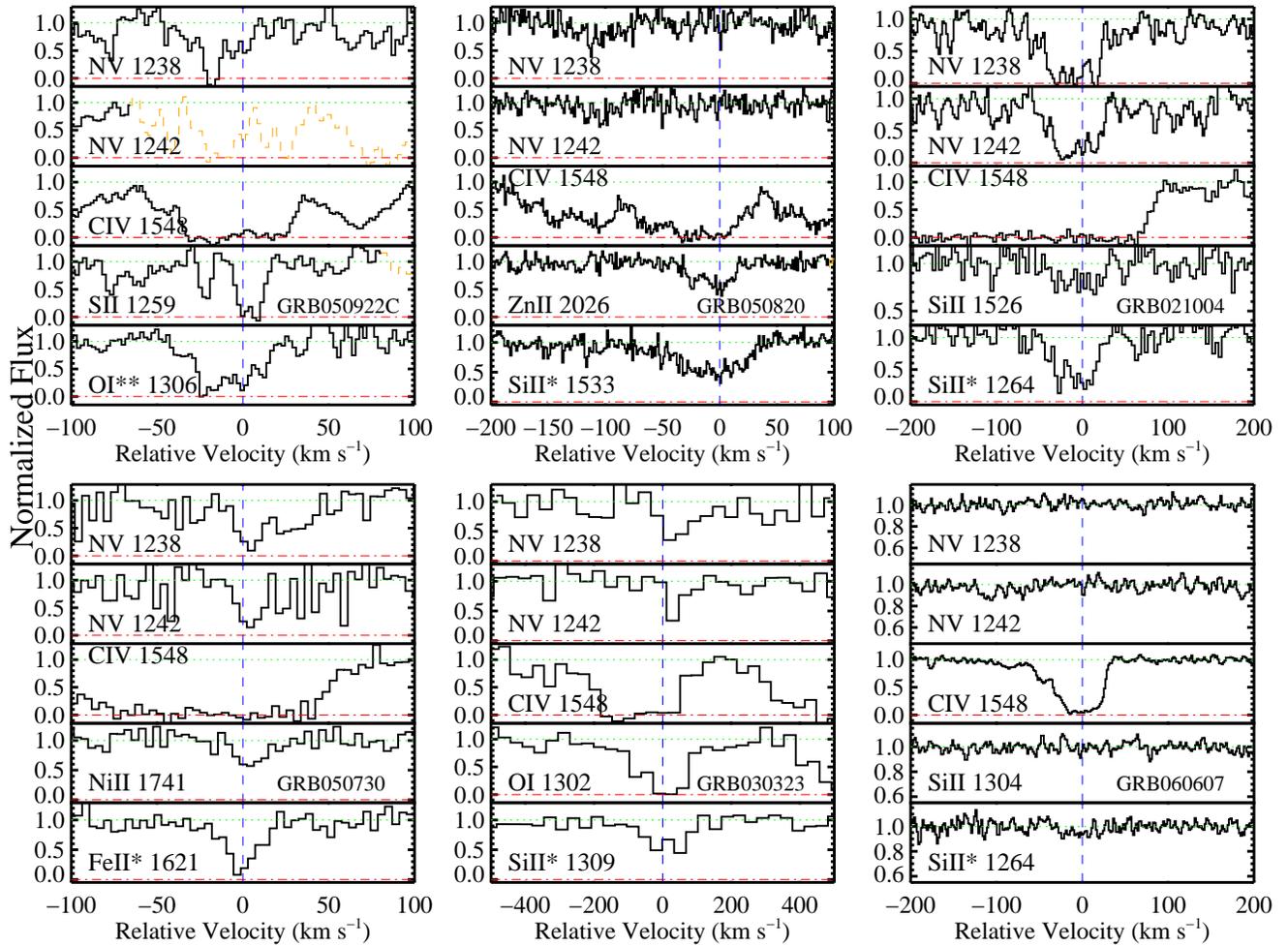}
\end{center}
%\plotone{f1.eps}
\caption{Velocity profiles of the \ion{N}{5} doublet (top two
panels in each sub-figure) compared against a low-ion resonance
and fine-structure transition from our sample of GRB afterglow
spectra with resolution $R > 2500$.  
We do not show the data for GRB~060206 (also included in the
analysis);  the interested reader should see \cite{twl+07}.
With the exception of GRB~060607,
the data reveal a positive detection of the \ion{N}{5} doublet.
Aside from GRB~050820, the \nion\ line-profiles are relatively
narrow and are roughly aligned with 
the low-ion resonance and fine-structure profiles.
}
\label{fig:data}
\end{figure*}

\section{A Survey for \ion{N}{5} Absorption in GRB Sightlines}

\subsection{The GRB Sample}

Our sample is comprised of all the 
published and/or publically available
GRB afterglow spectra that have resolution $R > 2500$ (FWHM~$<120\mkms$)
and coverage of the \lya\ profile and \ion{N}{5} doublet
from the ISM of the GRB host galaxy. 
This resolution criterion 
insures that the \ion{N}{5} doublet is well resolved and it also
establishes a sensitive detection limit: all of the datasets have a 
$1\sigma$ equivalent width limit of 30\,m\AA\ (or better) 
corresponding to $\N{N^{+4}} = 10^{13.15} \cm{-2}$. 
The sample is summarized in Table~\ref{tab:obs} and the
measured abundances are presented in Table~\ref{tab:summ}. 
Figure~\ref{fig:data} shows the corresponding line-profiles. 
In every case except GRB~060206, where we adopt the values reported
by \cite{twl+07}, 
we measured the rest-frame equivalent width of the
\ion{N}{5}~1238 transition and derived the \nion\ column density
using line-profile fitting techniques and/or the apparent
optical depth method \citep{savage91}.  In all but one case (GRB~060607),
we report the positive detection of an \ion{N}{5} doublet
within 100\kms\ of the peak optical depth of the
low-ion gas.  We associate this gas to the GRB host galaxy.

There are several common characteristics of the sample.
First, the peak optical depth and the integrated
column densities are large: five of the seven sightlines
exhibit saturated \ion{N}{5} absorption indicative of a column
density $\N{N^{+4}} \gtrsim 10^{14} \cm{-2}$.  
Second, these same cases have relatively narrow line-profiles
that are aligned or nearly aligned
with the peak optical depth of the low-ion profiles.
These characteristics are suggestive of a photoionized 
gas\footnote{Collisional ionization equilibrium would imply
$T \gtrsim 10^5$\,K and a Doppler parameter for the N gas of
$b \gtrsim 15 \mkms$.  Line-profile analysis of the echelle
spectra indicate Doppler widths less than this 
value \citep[e.g.][]{dpc+08}.}.
Third, the \ion{N}{5} profiles generally coincide in velocity
with the fine-structure lines.  By association,
we argue the majority of \nion\ gas is located within $\approx 1$\,kpc of
the GRB event \citep{pcb06}.  
Finally, there are no significant trends between the 
characteristics of the \nion\ gas and the ISM metallicity,
GRB redshift, or the \ion{H}{1} column density or the metal column
density of the ambient ISM.  

Two of the seven sightlines do not follow the general trends
described above.  The first (GRB060607) does not exhibit any
\ion{N}{5} absorption to very low limits.  This sightline is
unique for having a very low \ion{H}{1} column density and only
shows \ion{C}{4} and \ion{Si}{4} absorption at $z \approx z_{GRB}$.  
The other exceptional system (GRB050820) shows weak, broad \ion{N}{5}
absorption that is offset by $\delta v \approx -100 \mkms$
from the peak optical depth of the low-ions.  These characteristics
are fundamentally different compared to the remainder of the
sample and it suggests that more than one process is responsible
for the production of \ion{N}{5} gas along GRB sightlines.
In the following, however, we will focus on the majority of the
sample.

\subsection{Comparisons with other \ion{N}{5} Surveys}

%In several manners, the GRB sample is qualitatively different
%from comparable studies of galaxies at high $z$ along quasar sightlines
%(the damped \lya\ systems) and galaxies in the local universe.
The general characteristics of the GRB \ion{N}{5} sample contrast with 
those for \ion{N}{5} samples obtained from quasar sightlines through local
and high $z$ galaxies.
The most striking difference is 
the high detection rate of large \nion\ column densities in GRB sightlines. 
\cite{fpl+07} have presented the results of a search for 
\ion{N}{5} absorption in sightlines where they also surveyed
the O$^{+5}$ ion.  They reported only two positive \ion{N}{5} detections
(each with $\N{N^{+4}} \approx 10^{13.5} \cm{-2}$) among six
DLAs with both an \ion{O}{6} detection and \ion{N}{5} coverage.  They
set upper limits to the \nion\ column densities of $<10^{13}\cm{-2}$
in the remainder of cases.  
Similarly, an inspection of the public Keck/(HIRES+ESI) DLA database
\citep{pwh+07} indicates the incidence of \ion{N}{5} detections
to a column density limit of 10$^{13} \cm{-2}$ is less than 20\%
(see also Fox et al., in prep).
Therefore, the GRB
sightlines show more frequent and much stronger \ion{N}{5}
absorption than random sightlines through high $z$ galaxies.
Regarding the local universe, the majority of sightlines 
through the Galactic halo and disk do exhibit \ion{N}{5} 
absorption \citep{ssl97}, but only a few ($\approx 10\%$)
show column densities
$\N{N^{+4}} > 10^{14} \cm{-2}$ and none have 
$\N{N^{+4}} > 10^{14.5} \cm{-2}$ \citep{is04b}.  

Another important difference is that the \ion{N}{5} profiles
for GRB
are kinematically `cold': they are
narrow and have small offset (if any) from the low-ion
line-profiles.  This contrasts with the majority of Galactic detections
whose \ion{N}{5} line-profiles are systematically broader
than those for low-ions.  The Galactic \ion{N}{5} lines
have Doppler parameters \citep[$b > 30\mkms$;][]{ssl97}
that generally exceed those observed for the GRB data.  
By a similar token, very few Galactic \ion{N}{5}
profiles show peak optical depths $\tau_{1238} > 1$. 
In short, the 
profiles characteristic of GRB sightlines, i.e.\ narrow 
lines with large peak optical depths, are rarely observed
in any astrophysical environment of the local universe.
Regarding the \ion{N}{5} detections in high $z$ DLA galaxies,
the few examples with positive \ion{N}{5} detections are comprised
of multiple components whose widths more resemble the GRB profiles.
These examples, however, have much smaller peak optical depths 
than the majority of the GRB sample.
Furthermore, these lines do not trace the peak optical depths
of low-ion gas but are coincident with other high-ions 
(e.g.\ \ion{C}{4}, \ion{O}{6}) often with offsets of several
tens \kms\ from the low-ion features \citep{fpl+07}.
The only extragalactic environment where narrow, strong \ion{N}{5} profiles
have been observed is in gas associated with quasars \citep[e.g.][]{dcr+04}.  
These associated systems are modeled as photoionized material at
distances of a few tens kpc from the QSO.
In the following section, we will examine whether a similar process
-- photoionization by the GRB afterglow -- explains our observations.

\section{Discussion}

The observations presented in the previous section demonstrate
the nearly ubiquitous detection of \nion\ gas at $z\approx z_{GRB}$
in GRB afterglow spectra.  The \ion{N}{5} absorption lines generally 
(i) have large peak optical depths with large integrated column densities 
$\N{N^{+4}} \gtrsim 10^{14} \cm{-2}$; 
(ii) the profiles are relatively narrow ($<50\mkms$);
(iii) show small velocity offset
from fine-structure and resonant low-ion absorption; and 
(iv) exhibit no significant correlation with other
physical characteristics (e.g.\ metallicity) of the nearby ISM.
Similar to the fine-structure lines observed in GRB afterglow spectra
\citep{pcb06}, the prevalent detection
of strong \ion{N}{5} contrasts with sightlines through
galaxies in the local and high $z$ universe.  
Therefore, one infers that processes
related to the GRB, its progenitor, and/or its host galaxy
must produce the N$^{+4}$ gas.
We will now explore possible origins of this gas and its implications for
the GRB progenitor environment.  

\subsection{\ion{N}{5} Arising in the Halo and Neutral ISM Gas}

We begin by considering the galactic halo\footnote{Here, we use the
term galactic halo to describe gas that is distinct
from the neutral ISM of the galaxy, i.e.\  we (roughly) associate 
this component with any gas at scale-heights $\gtrsim 1$\,kpc from the
plane of the galaxy.} 
of the GRB host galaxy where 
one might expect a diffuse, hot baryonic component including \nion\ gas.
This expectation stems from surveys of the Galactic halo where one
observes a high covering fraction of high-ion
gas including \ion{O}{6} and \ion{N}{5} absorption \citep{sws+03,ssl97}. 
The relative column densities of these ions, however,
do not follow predictions from collisional ionization equilibrium (CIE) models
%Studies of the Galactic ISM indicate that \ion{N}{5} 
%absorption, like \ion{O}{6}, has a relatively high
%covering fraction in the Galactic halo
and researchers tend to invoke non-equilibrium scenarios
to explain the observations \citep[see][for a recent
review]{is04a}. 
These include turbulent mixing layers \citep{ssb93}, gas
shocked by supernovae remnants \citep{sm79}, and conductive interfaces 
\citep{sc92}.
None of these models describe all of the Galactic observations
and \cite{is04b} have concluded that several (if not all) of the
processes are likely to contribute.
It is reasonable to consider whether
these processes contribute to
the \ion{N}{5} absorption observed in GRB afterglow spectra.
Several arguments point against this hypothesis.
Regarding the theoretical models,
none of the mechanisms introduced to explain Galactic
N$^{+4}$ gas predict column densities significantly larger than
$10^{13}\cm{-2}$.  In part, this may be because the models
were computed to explain the Galactic data where one rarely finds
N$^{+4}$ column densities as large as the GRB sample.  Nevertheless, the
models cannot reproduce the large \ion{N}{5} optical depths unless
one presumes a very large number of these layers and/or remnants.
This `many absorber' scenario is challenged by the narrow 
\ion{N}{5} profiles observed for the majority of GRB sightlines 
(Figure~\ref{fig:data}), especially
if one allows that the N
abundance in GRB host galaxies is 
significantly sub-solar \citep{pcd+07}.

Furthermore, several lines of observational evidence challenge
interpreting the \nion\ gas as galactic halo material.  First, the GRB
\ion{N}{5} profiles have large optical depth, are more narrow, and are
more tightly correlated with the neutral ISM than the \ion{N}{5}
profiles observed in the Galactic halo.
%Instead, the \ion{N}{5} characteristics of GRB sightlines resemble the low
%and intermediate-ion profiles which trace photoionized or neutral gas.
Second, the coincidence of the \ion{N}{5} and fine-structure profiles
in velocity space suggests the \nion\ gas occurs within $\approx
1$\,kpc of the GRB.  This is an indirect argument because \nion\ does
not have its own fine-structure levels.  Nevertheless, the coincidence
in 4 of 5 sightlines showing \ion{N}{5} absorption is statistical
evidence that the \nion\ gas is located near the neutral phase.
Note, however, that 
the difference in ionization potential between these ions
means the gas cannot be precisely co-spatial.
Third, and perhaps most important, the data support the presence of
halo gas but not at the velocities of the \ion{N}{5} absorption.
\cite{pcw+08} have identified low column density features in strong
\ion{Si}{2} and \ion{Fe}{2} transitions that do not exhibit
corresponding \ion{Si}{2}$^*$ or \ion{Fe}{2}$^*$ absorption.
Therefore, these `clouds' lie at distances greater than a few kpc from
the GRB afterglow.  The clouds tend to show corresponding high-ion
absorption (\ion{Si}{4}, \ion{C}{4}) and have relative ionic column
densities suggestive of partially ionized gas.  \cite{pcw+08}
associate these clouds with the galactic halo.
This halo gas does not exhibit \ion{N}{5} absorption and we infer that
the clouds which do exhibit \nion\ gas are unrelated to the halo.
%We also note that significant offsets are observed between the
%\ion{N}{5} absorption and the peak optical depth of low-ions in
%QSO-DLA \citep{fox}.
Altogether, these theoretical and observational arguments disfavor a
galactic halo origin for the majority of \nion\ gas observed in GRB
sightlines.  The only obvious exception is the weak, broad \ion{N}{5}
doublet observed toward GRB~050820 which is also significantly offset
from the fine-structure absorption.  We await larger GRB samples to
assess the frequency of \ion{N}{5} detections like this one.

Granted the small velocity offset observed between the \ion{N}{5}
profiles and the low-ion transitions, one should consider whether the
gas is located near the neutral phase, i.e.\ within the ambient ISM of
the galaxy.  Indeed, observers have identified \nion\ gas in the
Galactic ISM and even our local bubble \citep{ssl97,wl05}.  Because
\nion\ is not produced by even bright O stars, it must trace
interstellar shocks from SN, X-ray emission from hot gas in the ISM,
or gas near white dwarfs \citep[e.g.][]{wl05,dr83}.  In comparison
with the GRB observations, however, the observed \nion\ column
densities of the Galactic ISM are generally small ($\lesssim 10^{13}
\cm{-2}$) and the line-profiles are significantly broader.  The values
are even extreme for \ion{O}{6} gas in the Galactic ISM \citep{bjt+08}
which always shows larger column densities than \ion{N}{5}
gas\footnote{We also note that the \ion{O}{6} lines with largest
  column densities have the largest Doppler parameters.}.  One
counter-example is observed along the sightline to H1821+643
\citep{ssl95}.  In this case, one observes a strong \ion{N}{5} profile
that is likely associated with the planetary nebula K1-16.  At the
redshifts of GRB host galaxies, however, planetary nebulae are rare or
non-existent and we expect this process to have a negligible
consequence.  In summary, although we cannot unambiguously rule out
the \ion{N}{5} gas arising in the ambient ISM, we consider this to be
an unlikely scenario.
%Similarly, one can rule out photoionization by the GRB or its
%progenitor for an ISM scenario because the gas lies at too great
%of a distance ($>100$\,pc).  Furthermore, only an extremely
%hard radiation field could yield a detectable column density of
%\nion\ without photoionizing all of the low-ion gas.

\subsection{\ion{N}{5} Associated with the Starburst Phenomenon}

Empirically, one
observes \ion{N}{5} absorption in the individual and integrated
spectra of $z \sim 3$ star-burst galaxies \citep[Lyman break galaxies,
  LBGs][]{prs+02,shapley03}.  The VLT/UVES spectrum of the lensed cb58
galaxy \citep{spp02}, for example, shows a strong \ion{N}{5} profile
at a velocity consistent with other high-ion and low-ion (resonant and
fine-structure) lines observed in the spectrum.  All of this gas is
offset by $\approx -200\mkms$ from the observed nebular emission lines
indicating a galactic-scale outflow, presumably associated with the
current burst of star-formation.  The physical origin of \nion\ has
not been established for these high $z$, star-burst galaxies.
%Because \nion\ is difficult to produce via photoionization with
%stellar sources, we speculate that it is induced by shocks in
%the galactic-scale outflow, e.g.\ via collisional ionization or
%by x-ray photons from hot gas.  
Nevertheless, if GRB host galaxies are also driving galactic-scale
winds \citep[e.g.][]{tgs+07,pcw+08}, then it is possible that the
observed \ion{N}{5} absorption is related to this phenomenon.  While
these mechanisms deserve further attention, we note that the majority
of GRB host galaxies are not massive star-bursts but more resemble
galaxies like NGC~1705, a local, post star-burst dwarf galaxy
\citep{cpb07}.  While a galaxy like NGC~1705 also may exhibit an
outflow \citep{hl97}, the observed P-Cygni \ion{N}{5} absorption is
not associated with the galactic-scale outflow.  In this respect, the
\ion{N}{5} absorption associated with bright LBGs may not be relevant
to the GRB afterglow spectrum.

\begin{figure}
%\epsscale{0.8}
%\plotone{../Figures/fig_ions.ps}
\centerline{\epsfxsize=\hsize{\epsfbox{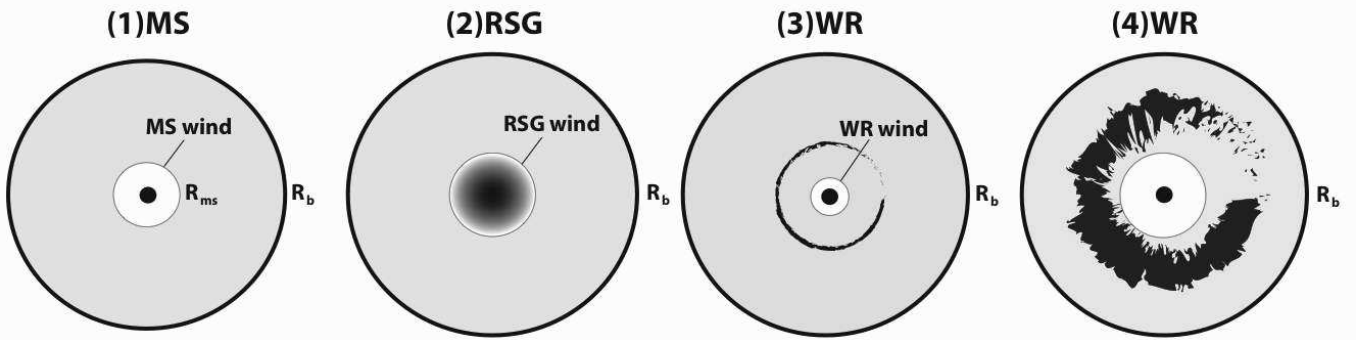}}}
%\plotone{f2.eps}
\caption{
A strong wind from a massive star sweeps up the ambient gas forming a
stellar wind bubble.  Stellar wind bubbles have a double shock
structure. An outer shock sweeps up the interstellar medium and forms
a thin, radiatively cooled shell of swept-up gas by a contact
discontinuity. Since massive stars pass through phases with different
mass-loss rates during their evolution, there will be a variety of
structures in the surrounding gas (Garcia-Segura et al. 1996). The
first, a rarefied, main-sequence wind sweeps up the ambient ISM (1).
A dense, slow wind that expands into the rarefied interior of the
main-sequence bubble follows as the star evolves through the RSG phase
(2). The shell swept up from the main-sequence wind at $R_{\rm b}$
remains almost constant during this time. This RSG wind forms a dense,
high metallicity circumstellar medium surrounding the star. Finally, a
fast wind from the WR phase (3) sweeps up this previous slow wind,
forming a ring WR nebula. As the WR shell is decelerated by the RSG
shell, dense clumps (which have higher inertia) overrun the RSG wind
material. After that, the whole interior bubble breaks out and the hot
shocked WR wind expands into the remnant main sequence bubble (4).
}
\label{fig:winds}
\end{figure}

\subsection{Do Stellar Winds Produce \ion{N}{5} gas?}

Massive stars are observed to generate stellar winds during the course
of their main-sequence lifetimes and frequently during the stellar
phases that follow.  These stars are subject to severe mass-loss
throughout their lives, e.g. Galactic stars with initial masses above
$\sim 25\msol$ are thought to lose more than half of their initial
mass before their final supernova explosions. 
Massive stars may pass through several phases
having very different stellar winds during their evolution, which may
produce a wide variety of structures in the surrounding gas. These
stellar winds drive supersonic shock waves into the ambient gas,
sweeping and heating it up. Under the presumption that GRB progenitors
are massive stars, it is feasible that shocks produced by their
stellar wind would produce an observable quantity of \ion{N}{5} gas.

The detailed dynamical evolution of the circumburst medium around
massive stars is complex. Since massive stars pass through phases with
very different mass-loss histories during their evolution, there will
be a variety of structures in the surrounding gas. Indeed, at solar
metallicities main sequence stars of moderate mass ($25-40\msol$) are
thought to develop into red supergiants (RSG) and thereafter
Wolf-Rayet (WR) stars. The winds of possible GRB progenitors have been
explored by a number of authors
\citep{glm96,rgs+05,vlg05,vly+08,egd+06}. These progenitor stars
are thought to be the highly evolved descendants of main-sequence
stars with initial masses larger than about 20 $\msol$.

In these models, there are up to three consecutive types of winds (see
Figure~\ref{fig:winds}). 
The first is a rarefied, fast wind from the main sequence (MS)
star that sweeps up the ambient ISM, forming a main-sequence stellar
wind bubble. Typical velocities are $v_{\rm ms} \sim 10^3$ km/s, and
mass-loss rates are $\dot{M}\sim 10^{-6}-10^{-7}\;\msol$/yr, with
lifetimes of $\tau_{\rm ms}\sim 10^6$ yr \citep{hkv+92}.
A dense and slow wind follows as the star evolves to a RSG phase. This
wind expands into the rarefied interior of the main-sequence
bubble. This wind forms a dense, high-metallicity circumstellar medium
surrounding the star. RSG wind strengths are $v_{\rm rsg} \sim 10-25$
km s$^{-1}$ and $\dot{M}\sim 10^{-4}-10^{-5}\;\msol\;{\rm yr}^{-1}$,
with lifetimes of $\tau_{\rm rsg}\sim 10^5$ yr
\citep{humph91}.
Finally, a fast wind from the WR phase with $v_{\rm wr} \sim 10^3$
km/s and $\dot{M}\sim 10^{-5}-10^{-6}\;\msol\;{\rm yr}^{-1}$
\citep{willis91},
sweeps up the previous slow wind just prior to the death of the
star. When the fast WR wind stars blowing, it sweeps up the RSG wind
material into a shell, forming the observed WR ring nebula.  Finally,
the WR ring nebula breaks out of the RSG wind. After that, the hot
shocked, WR wind expands into the remnant main sequence bubble.

During the main-sequence phase, the stellar mass-loss increases
gradually, while the wind velocity decreases slightly, which results
in an almost constant mechanical luminosity
\begin{equation}
L_{\rm w}={1 \over 2}\dot{M}_{\rm w}v^2_{\rm w}.
\end{equation}
This behavior allows us to easily compute the time evolution of the
main-sequence bubble using the analytical solutions given by
\citet{wmc+77}. Their adiabatic thin-shell solutions are summarized
as follows:
\begin{eqnarray}
E_{\rm th}={5 \over 11}L_w\;t,\\ P_{\rm b}={7 \over
  (3850\pi)^{2/5}}L_w^{2/5}\rho_0^{3/5}t^{-4/5},\\ R_{\rm b}=\left({125
  \over 154\pi}\right)^{1/5}L_{\rm w}^{1/5}\rho_0^{-1/5}t^{3/5},
\end{eqnarray}
where $E_{\rm th}$ is the thermal energy of the hot shocked
main-sequence gas, $P_{\rm b}$ is thermal pressure of the shocked gas,
$R_{\rm b}$ the outer radius and $\rho_o$ is the ISM constant
density. These equations assume that the forward shock is completely
radiative, whereas the hot bubble of the shocked wind material is
adiabatic, which is a good approximation given that the hot shocked
gas radiates poorly. The resulting shocked temperature can thus be
estimated by
\begin{equation}
T_{\rm shocked}={3 \over 16}{\mu m_H \over k}\Delta v_w^2 \sim 10^{5}\left({\Delta v_w \over 10^2\;{\rm km\;s^{-1}}}\right)^2\;{\rm K}.
\label{shockt}
\end{equation}
In addition, the above formalism assumes that the ISM is cold, so that
it has no significant thermal pressure.

The approximation of a thin shell thus gives an estimate for the size
of the circumstellar main-sequence bubble
\begin{equation}
R_{\rm b}=52.9 \left({L_{\rm ms} \over 10^{36}\;{\rm
    erg\;s^{-1}}}\right)^{1/5}\left({\rho_0 \over 10^{-23}{\rm
    \;g\;cm^{-3}}}\right)^{-1/5}\left({t_{\rm ms} \over 5 \times
    10^6\;{\rm yr}}\right)^{3/5}\;{\rm pc}.
\label{bubblesize}
\end{equation}
The wind termination shock at the end of the main-sequence phase is
located at a radius $R_{\rm ms}$ such that the thermal pressure in the
shocked wind material equals the hot bubble pressure. The
post-termination shock pressure is roughly equal to the ram pressure
in the wind, $P_{\rm ram}=\rho_{\rm ms} v_{\rm ms}^2$, so that
\begin{equation}
R_{\rm ms}=7.8\;\left({\dot{M}_{\rm ms} \over 10^{-6}\;\msol\;{\rm
    yr^{-1}}}\right)^{1/2}\left({v_{\rm ms} \over 10^3\;{\rm
    km\;s^{-1}}}\right)^{1/2}\left({P_{\rm b} \over 3 \times
  10^{-12}\;{\rm dyne\;cm^{-2}}}\right)^{-1/2}\;{\rm pc}.
\label{shockmssize}
\end{equation}
Eventually, the star leaves the main sequence and passes through the
RSG phase. In the transition phase from main-sequence to RSG stage,
the stellar wind increases its density and decreases its velocity. In
total, the ram pressure is reduced 
(i.e.\ $\rho_{\rm ms} v^2_{\rm ms} \ge \rho_{\rm rsg} v^2_{\rm rsg}$), 
and the wind
terminal shock loses its equilibrium position. It collapses and finds
a new stationary location on the hydrodynamic timescale of the hot
bubble.  Once a new equilibrium point is found, a shell of shocked RSG
wind stars to build up (although the RSG wind is slow, it is
supersonic given its low temperature which results in an extremely low
sound speed). After the RSG phase, the star evolves directly to the WR
phase. The fast wind sweeps up the previous slow RSG wind material
into a shell. $L_{wr}$ exceeds the mechanical luminosity of the RSG
wind by several orders of magnitude and as a result, the WR-shell will
eventually break out of the RSG wind. After that, the hot shocked, WR
wind expands into the remnant main-sequence bubble.

To produce a detectable \ion{N}{5} feature, the post-shock temperature
must be $T_s \approx 10^5$\,K to collisionally ionize nitrogen to
\nion\ and the shock must extend beyond $R \approx 20$\,pc to avoid
photoionization by the GRB afterglow (see the following section).
These constraints rule out an RSG wind as this is too weak to extend
beyond a few pc. What is more, the RSG wind is ultimately heated and
accelerated to large velocities by the WR wind before core collapse.
Meanwhile, the speeds of MS and WR winds lead to a shock temperature
$T_s > 10^6$\,K \citep[e.g.][]{vly+08}.  The only exception is in the
dense shell of swept-up material at the wind's edge but this rapidly
cools to $\approx 10^4$\,K.  Altogether, we conclude that stellar winds
associated with the GRB progenitor will not lead to significant
\ion{N}{5} absorption.

In lieu of shocked gas from a single stellar wind, one could invoke
shocks related to an expanding `superbubble', perhaps associated with
a starburst \citep[e.g.][]{tbr87}.  
Following equation~\ref{shockt}, the appropriate
temperature for \ion{N}{5} is achieved for an expansion speed of
$\approx 100 \mkms$.  This simple scenario
yields the following predictions for a shock-heated \ion{N}{5} gas:
the \ion{N}{5} absorption will be offset by $\delta v \approx
-100\mkms$ from the ISM gas and have a Doppler parameter consistent
with $T \gtrsim 10^5$\,K (i.e.\ $b \gtrsim 15 \mkms$). Reviewing the
GRB sightlines comprising our sample, GRB~050820 shows a surprisingly
good match to these characteristics.  Although other processes could
explain the \ion{N}{5} gas associated with this GRB (e.g.\ diffuse gas
in the galactic halo), one may associate this \ion{N}{5} detection
with shocked gas.  Conversely, the remainder of \ion{N}{5} detections
have small velocity offsets from the ISM and do not exhibit broad
line-profiles characteristic of $T \approx 10^5$\,K gas.  For the
majority of the GRB sample, therefore, we must consider an alternative
process.  Furthermore, the absence of \ion{N}{5} detections at $\delta
v \ll 0\mkms$ in these sightlines constrains the nature of stellar
winds and the ISM.  The data suggest that the stellar wind has a
termination radius less than 10\,pc such that the afterglow has
photoionized this material.  We will explore this latter idea at
greater length in the next section.

\subsection{Photoionization by the Afterglow}

In the previous sections we argued that the halo and neutral ISM of
the host galaxy as well as material shock-heated by the progenitor's
stellar wind are unlikely to yield \nion\ gas with characteristics
resembling the majority of our GRB \ion{N}{5} sample.
%one is driven toward the local environment of the GRB, 
%i.e.\ the surrounding \ion{H}{2} and/or 
%circumburst region of the progenitor.  
Motivated by the narrow \ion{N}{5} line-profiles, we turn to a model
where the \nion\ gas arises from photoionization.  This cannot include
ionization by an OB association because these stars emit far too few
photons at $h \nu > 77$\,eV to produce a meaningful column density of
\nion.  Even a 40\,$M_\odot$ star emits too few photons during its
lifetime to produce a significant \nion\ column density.  The
progenitor stars of GRBs, however, may be very massive stars that
undergo a Wolf-Rayet phase with effective temperatures exceeding
100,000\,K \citep{hmm05,hfs+06}\footnote{Note that although rapidly
  rotating models suggest a higher flux of $h\nu \approx 100$\,eV
  photons (Meynet, priv.\ communication), this will not change our
  conclusion.}.  These could produced measurable column densities of
\nion, but only at distances $r \ll 1$\,pc from the WR star.  We now
demonstrate that this gas is photoionized by the GRB afterglow.

Owing to synchrotron processes, GRB afterglows have a roughly
power-law spectrum with emission extending from the optical to X-ray
frequencies.  The afterglow, therefore, will initiate an ionization
front giving highly ionized material at small radii ($r < $1pc) and
progressively less ionized gas at larger radii as the photon flux
decreases by $r^{-2}$.  The photon flux is sufficiently high that the
$r^{-2}$ dependence dominates over any attenuation (shielding) by the
surrounding medium.  One can estimate the distance where a specific
ion (e.g.\ \nion) will reach a maximum ionization fraction by
comparing the photon flux $f_\gamma$ which ionizes the next lower
ionization state (i.e.\ \nthr) against the photoionization
cross-section of that ion (i.e.\ $\sigma({\rm N^{+3}})$).  For
$f_\gamma \sigma \ll 1$, there will be a negligible ionization
fraction.  For $f_\gamma \sigma \gg 1$, it is likely that the ion will
be fully ionized to a higher state.  The peak in the ionization
fraction is therefore likely to occur where $f_\gamma \sigma \sim 1$.

We can estimate this distance by adopting the afterglow of GRB~050730
which we parameterize as
\begin{equation}
\L_\nu = 7.4\sci{30} \; \ltp \frac{h\nu}{80\,{\rm eV}} \rtp^{-1.8}
\ltp \frac{t_{obs}}{100\,{\rm s}} \rtp^{-0.3} \;\; {\rm erg \; s^{-1} \;
Hz^{-1} \;\;\;\; .}
\label{eqn:after}
\end{equation}
Here, the frequency $\nu$ corresponds to the rest-frame and $t_{obs}$
is time in the observer frame.  We integrate from $t_{obs} = +10$s
(i.e.\ we ignore effects associated with the prompt phase) to $t_{obs}
= +1000$s which is a time typical for the onset of high-resolution
spectroscopic observations.  We calculate $\phi_\gamma = 3\sci{58}$
photons emitted with energies in the interval $77\,{\rm eV} < h \nu <
95\,{\rm eV}$ which bounds the ionization potentials of \nthr\ and
\nion.  At 1pc, this implies a photon flux of $f_\gamma \equiv
\phi_\gamma/(4\pi r^2) = 2.4 \sci{20} \cm{-2}$ which exceeds the
column density of N nuclei for $n_H < 10^5 \cm{-3}$ assuming a solar
abundance (i.e.\ shielding is likely negligible).  Finally, one
compares the photon flux with the cross-section of \nthr, $\sigma({\rm
  N^{+3}})$.  Using the parameterization of \cite{verner96},
$\sigma({\rm N^{+3}}) = 10^{-18} \cm{2}$ at $h\nu = 80$eV.
%At 1pc, the fraction of \nthr\ gas at
%$t = +1000$s is negligible:  
%${\rm N^{+3}_f/N^{+3}_i} \approx \exp[-f_\gamma 
%\sigma({\rm N^{+3}}) \approx \exp[-200]$.   
The distance where \nion\ should reach a maximum value is roughly
where the photon flux matches the cross-section, i.e.,
\begin{equation}
r_{peak} \approx \ltp \frac{\phi_\gamma \snt}{4\pi} \rtp^{1/2}
\end{equation}
or $\approx 15$\,pc for our example.

We note that the spectral slope assumed here is steeper than
most other afterglows and therefore implies a lower integrated
photon luminosity at 100\,eV than may be typical (some
afterglows emit in excess of 10$^{61}$ photons
in the 77 to 95\,eV interval).
Furthermore, we have ignored the prompt emission of \nion\ ionizing
photons. 
Adopting the spectral parameterization of the prompt X-ray emission
observed for GRB~060614 as an example \citep{bk07},
we estimate $10^{58}$~photons emitted during the prompt phase 
$t_{obs} \le 100$s.
This low redshift burst has a smaller X-ray luminosity than typical
and we estimate the prompt phase can contribute in excess of 
$10^{60}$photons.   If $E_{peak}$ evolves like
$t_{obs}^{-2}$, then to first order the prompt and afterglow phases
will have comparable time-integrated photon fluxes at 100\,eV.

To verify the distance estimate above
and to explore a range of physical conditions, we
have performed a series of time-dependent, photoionization
calculations for the propagation of ionizing flux through a constant
density medium \citep[e.g.][]{pl02}.  Our calculation follows the
photons through a series of constant density, optically thin layers
allowing for absorption by H, He, C, N, and O.  We ignore
recombinations because the timescales exceed (by orders of magnitude)
the duration of the observations.  In the following, we have adopted
the afterglow of GRB~050730 integrating from $t_{obs} = +10$s to
+1000s.  In this respect, the calculation yields the ionization state
of the gas that photons emitted at $t_{obs} = +1000\,{\rm s}+\delta t$
would `observe' on their travel to Earth.

\begin{figure}
\begin{center}
\includegraphics[height=6.5in,angle=90]{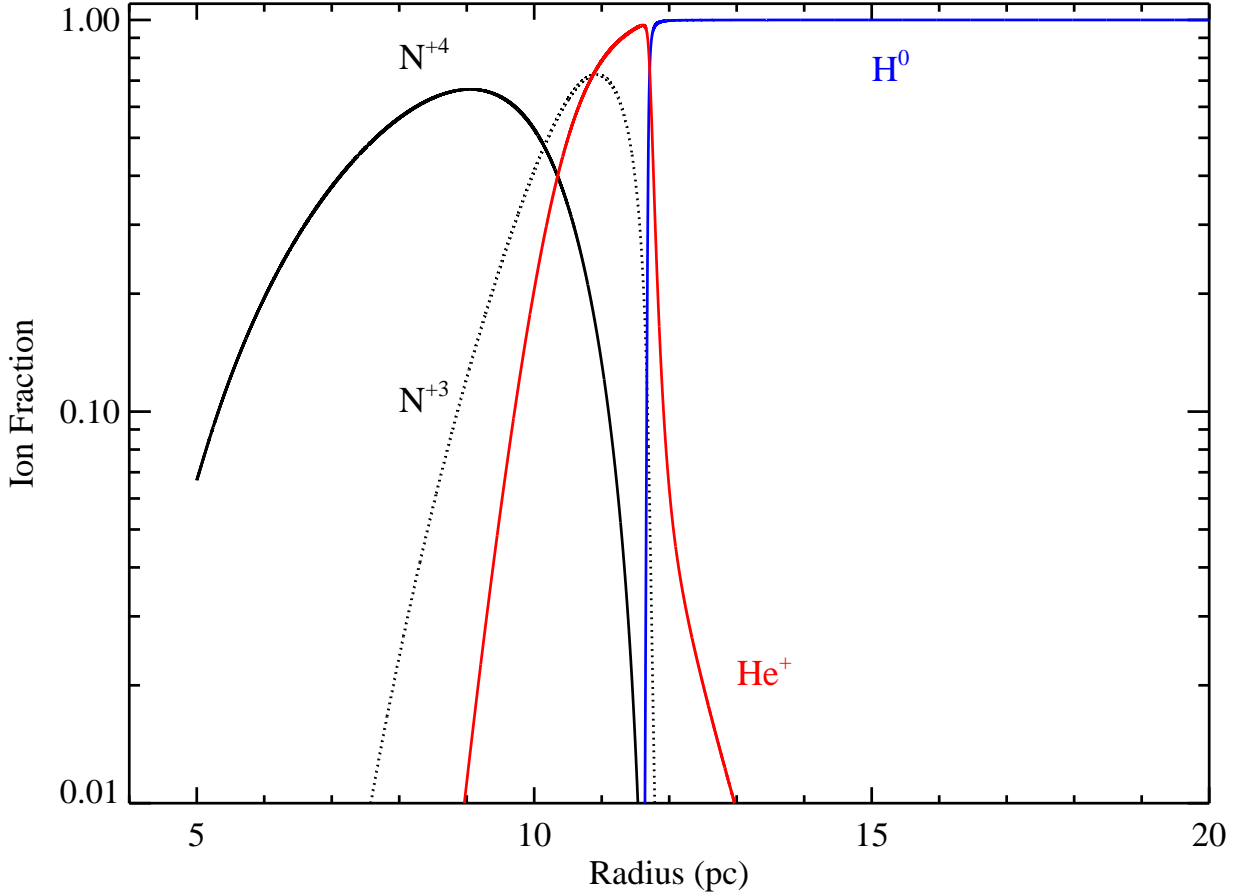}
\end{center}
\caption{
Ion fractions for a series of ions (i.e.\ X$_i$/X) as a function
of distance from the GRB afterglow.
The calculation assumes a constant density medium ($n_H = 10 \cm{-3}$)
and the total photon flux from an afterglow with 
$L_\nu = $ integrated from $t= +10$s to +1000s in the observer
frame.  Therefore, the calculations represent the ion fractions
that a photon emitted at $t = +1000\,{\rm s} + \delta t$
would `observe' as it traveled through the medium.
}
\label{fig:ions}
\end{figure}

\begin{figure}
%\epsscale{0.8}
%\plotone{f4.eps}
%\plotone{../Figures/fig_rnevol.ps}
\begin{center}
\includegraphics[height=6.5in,angle=90]{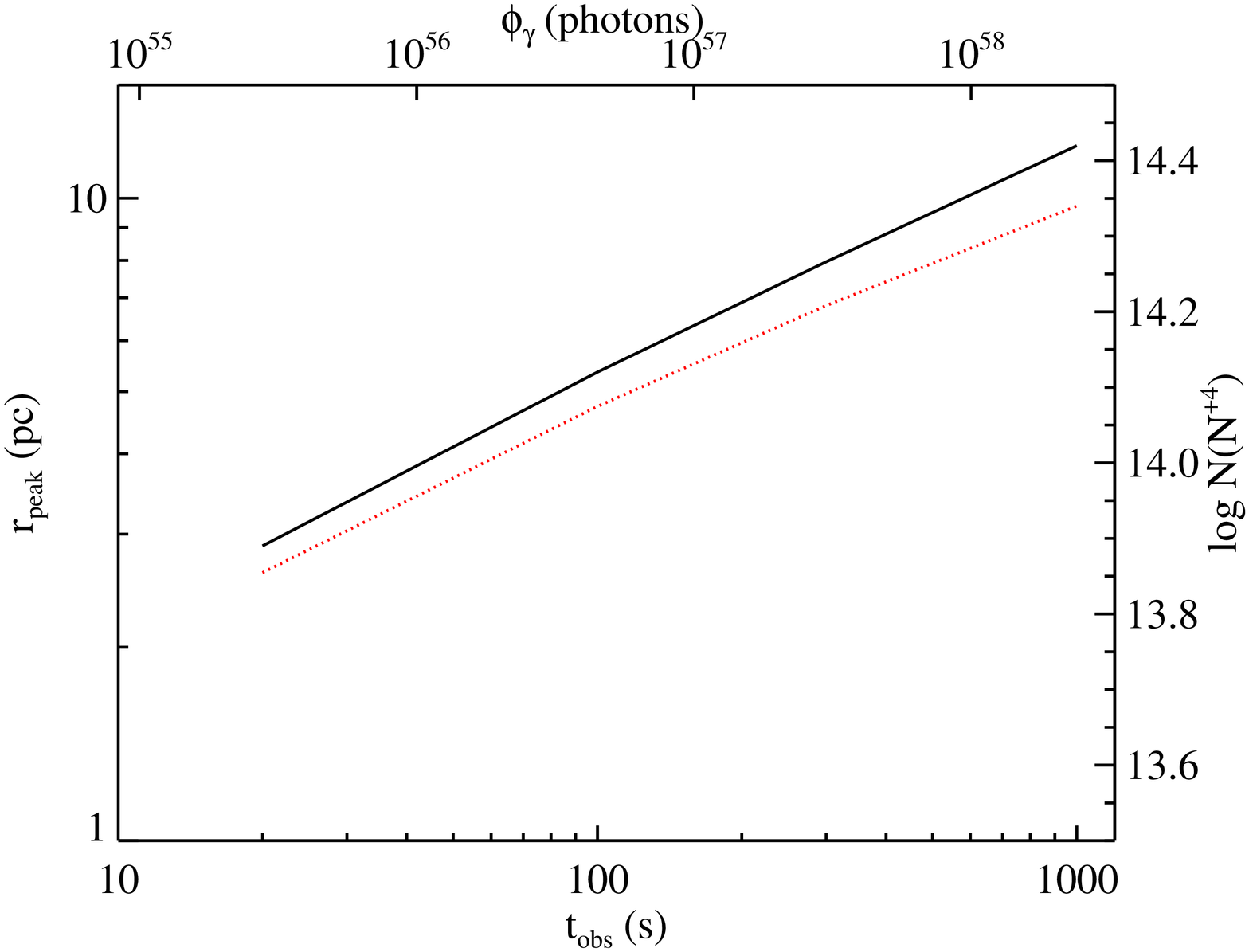}
\end{center}
\caption{
(solid black line):
Evolution in the distance from the afterglow
where the ionization fraction of \nion\ peaks
as a function of the time (x-axis) or number of 
ionizing photons (y-axis) emitted.
(dotted red line):
Evolution in the column density of \nion\ 
as a function of the time (x-axis) or number of 
ionizing photons (y-axis) emitted.
Both calculations assume the afterglow parameterization
given by equation~\ref{eqn:after}, $n_H = 1 \cm{-3}$,
and [N/H]~$= -1$\,dex.
}
\label{fig:time}
\end{figure}

The curves in Figure~\ref{fig:ions} show the ionization fractions of
N$^{+3}$, \nion, He$^+$, and H$^0$ as a function of distance from the
GRB afterglow for $n_H = 10 \cm{-3}$.  We find that the majority of
nitrogen gas at $r_{peak} \approx 10$\,pc has been photoionized to
\nion, while nitrogen gas at smaller distances is in higher ionization
states and nitrogen gas at greater distances are in lower ionization
states. The total \nion\ column density predicted from this
calculation (assuming $n_H = 10 \cm{-3}$) is

\begin{equation}
\N{N^{+4}} = 10^{14} \cm{-2} \; 
\ltk {\rm \frac{(N/H)}{10^{-6}}} \rtk 
\;\;\;\;  [n_H = 10 \cm{-3}],
\label{eqn:nv}
\end{equation}
where we note that (N/H)$= 10^{-6}$ corresponds to 1/100 solar
abundance.  This calculation shows that even a modest density,
sub-solar gas can produce an \nion\ column density consistent with the
observations.  We have repeated the calculations for a range of $n_H$
values and find the \nion\ column density scales with $n_H$ for small
$n_H$ values but has an approximately $n_H^{1/2}$ dependence for $n_H
> 5 \cm{-3}$.  The calculations also indicate that the \nion\ ionic
fraction peaks at a slightly smaller radius (by a few pc) for larger
$n_H$ but roughly the same position for smaller $n_H$ values.  These
results are sensitive, however, to assumptions on the ionization state
of the gas prior to the afterglow.  Here, we have assumed the gas is
neutral at $t_{obs}=10$s at all radii $r > 1$pc.  If one allows for a
pre-existing \ion{H}{2} region extending to greater than 30\,pc
\citep[e.g.][]{wph+08}, this gives more \nion\ gas at larger radii and a
larger total \nion\ column density due to reduced shielding by
\ion{H}{1} gas but the differences are modest ($<30\%$).  We have
also explored afterglows with a range of luminosities.  This is
identical to studying a single system at various times $t_{obs}$
because the only relevant quantity is the integrated photon flux prior
to the time of observation.  Not surprisingly, both $r_{peak}$ and the
integrated \nion\ column densities increase with the afterglow
luminosity (Figure~\ref{fig:time}).

The results presented in Figure~\ref{fig:ions} and described above
reveal a generic prediction for GRB afterglows.  Just as the $\approx
6$\,eV far-UV photons excite Si$^+$ and Fe$^+$ ions out to 1\,kpc
distance, the $\approx 80$\,eV photons from the afterglow photoionize
nearly all nitrogen at $r \approx 10$\,pc to \nion.  
Furthermore, the strengths of the \ion{N}{5} absorption should
not be correlated with \ion{H}{1} column density and will
at best be loosely correlated with the low-ion column density.
These predications are consistent with the current sample
of observations.   The photoionization scenario described
here will also imply significant absorption by other high-ions, e.g.
\ion{C}{4}, \ion{O}{6} transitions.  Figure~\ref{fig:data} shows that we
observed strong \ion{C}{4} absorption at the velocity of 
\ion{N}{5} in every case.  We find, however, that the signal-to-noise
and line-blending in the current spectra prohibit a meaningful search
for the \ion{O}{6} doublet.
In summary, we
predict detectable \ion{N}{5} absorption in all GRB afterglow spectra
except those where the gas at $r \approx 10$pc has very low nitrogen
density $n_N \equiv n_H ({\rm N/H})$ or has been collisionally ionized
to higher states (i.e.\ $T > 10^6$\,K).

A high temperature ($T > 10^6$\,K; shocked wind) and low
density gas is in fact predicted at $r \approx 10$\,pc in the stellar winds of
GRB progenitors \cite[e.g.][]{vlg05}.  If we are to explain narrow
\ion{N}{5} absorption with small velocity offset from the neutral gas,
one must conclude that the stellar wind does not extend beyond $r
\approx 10$\,pc.  Referring to equation~\ref{bubblesize}, this implies
\begin{equation}
\rho_0 \gtrsim 8 \times 10^{-20} \left({L_{\rm ms} \over 10^{36}\;{\rm
    erg\;s^{-1}}}\right)\left({t_{\rm ms} \over 5 \times 10^6\;{\rm
    yr}}\right)^{3}\;{\rm \;g\;cm^{-3}}.
\label{eqn:rholim}
\end{equation}
This restriction is, however, only valid if the ISM is cold so that
the circumstellar bubble expands supersonically with respect to the
surrounding ISM. For a sufficiently high ISM temperature, the thermal
pressure of the ISM will act as an extra confining force. An example
of this is an \ion{H}{2} region. In such an environment, the main
sequence wind can never create a supersonically expanding
shell. Pressure balance between the thermal pressure in the hot wind
bubble and the thermal pressure of the ISM (with $n_0$ particle
density)
\begin{equation}
P_{0}= 10^{-10} \left({n_{0} \over 1\;{\rm cm^{-3}}}\right) \left({T
  \over 10^6\;{\rm K}}\right)\;{\rm dyne\;cm^{-2}},
\end{equation}
gives the radius of the wind termination shock
\begin{equation}
R_{\rm ms}=1.2\left({\dot{M}_{\rm ms} \over 10^{-6}\;\msol\;{\rm
    yr^{-1}}}\right)^{1/2} \left({v_{\rm ms} \over 10^3\;{\rm
    km\;s^{-1}}}\right)^{1/2}\left({P_0 \over 10^{-10}\;{\rm
    dyne\;cm^{-2}}}\right)^{-1/2}\;{\rm pc}.
\end{equation}
Another scenario that would avoid the complications due to a steady
stellar wind is a progenitor system that moves rapidly through the
interstellar medium \citep[e.g.][]{hfs+06,vla+06}. 
A complimentary conclusion can be drawn from this
analysis: GRB progenitor models that predict stellar winds extending
beyond 10\,pc must invoke a new process to explain the strong, narrow
\ion{N}{5} profiles with small offset observed for the majority of GRB
sightlines.

These issues suggest an additional problem: Where is the gas related
to afterglow photoionization in GRB~050820 for which one only observes
relatively weak, broad \ion{N}{5} absorption offset by $\delta v
\approx -100\mkms$ from the low-ion gas?  In the previous section, we
suggested that the \ion{N}{5} gas for GRB~050820 may be explained by
shock-heated gas associated with an expanding superbubble.  Provided
this bubble extends beyond 10\,pc, the material photoionized by the
afterglow will occur at $\delta v < 0 \mkms$.  Furthermore, this gas
is likely to have temperatures that ionize N beyond \nion.

Adopting the afterglow photoionization model, we can use the observed
\nion\ column density to constrain $n_N$ at $r \approx 10$pc from the
afterglow.  Let us consider a few examples from the current sample
under the assumption that the afterglow ionization scenario is the
only mechanism relevant to \nion\ production.  For those sightlines
with $\N{N^{+4}} \gtrsim 10^{14} \cm{-2}$, we derive this constraint:
$\log (n_H/\cm{-3}) + [{\rm N/H}] \gtrsim -1$\,dex.  This is a
relatively modest density and enrichment level; if GRB progenitors
arise in star-forming regions, one may expect these conditions to be
satisfied for nearly every GRB.  The non-detection of \nion\ gas
toward GRB~060607, therefore, has a surprising implication.  The upper
limit of $\N{N^{+4}} < 10^{12.6} \cm{-2}$ gives $\log (n_H/\cm{-3}) +
[{\rm N/H}] < -2.7$\,dex.  This observation suggests both a metal-poor
gas and a low-density medium surrounding the GRB.  It is interesting
to note that this GRB also shows the lowest \ion{H}{1} column density
of any GRB sightline to date \citep{cpg07}.  This may indicate that
some GRB events occur outside of both their star-forming regions and
the ISM of the host galaxy.

If one allows that the metallicity of the neutral ISM (inferred from
low-ion transitions) is applicable for the gas near the GRB, then one
can constrain the $n_H$ density alone.  In a few cases (050730,
050820, 050922C), one observes \ion{N}{1} transitions and can
constrain N/H directly \citep{pcd+07}.
%If one assumes these metallicities
%apply to at $r \approx 10$pc from the afterglow, then the measured \nion\
%column density gives a direct constraint on $n_H$.  
For example, the afterglow spectrum of GRB~050922C shows [N/H]~$< -4$
and $\N{N^{+4}} > 10^{14.2} \cm{-2}$.  This implies $n_H > 10^3
\cm{-3}$ unless one assumes the gas is enriched in N local to the GRB
\citep{dpc+08}.  We note that this density is comparable to the value
needed to confine a stellar wind to $r < 10$\,pc (see above).  A
summary of the constraints on the physical conditions for our dataset
is given in Table~\ref{tab:phys}.

%A final implication of the afterglow ionization scenario is that the
% data provide a snapshot of the velocity field at $r \approx 10$\,pc
% along the sightline.  This can be compared with the velocities of
% the neutral gas and, in principle, nebular emission lines
% \citep[e.g.][]{tgs+07}.  With the exception of GRB~050820, the
% observations indicate modest differential motions between the
% \ion{H}{2} region and the neutral ISM, i.e.\ one rules out a
% fast-moving wind or shock.  For GRB~050820, one observes \nion\ gas
% offset by $\delta v \approx -100 \mkms$ from the peak optical depth
% of the low-ions.  [comment further?]
Ultimately, the most valuable constrains may come from a time-series
of spectroscopy \citep[e.g.][]{vls+07} to investigate variability in
the kinematics.  Figure~\ref{fig:time} indicates that $r_{peak}$
increases by factors of a few during the first 1000s of the afterglow.
Therefore, one would be sensitive to the kinematics of the gas at a
range of radii allowing constraints on the differential motions at
these distances.
We have searched for temporal variations in the strength and
velocity of the \ion{N}{5} gas in the Keck/HIRES spectrum
of GRB~050820 (two exposures of 900s starting at $t_{obs} \approx 3300$s)
and the Magellan/MIKE spectrum of GRB~050730
(three 1800s exposures starting at $t_{obs} \approx 4$hr)
\citep{pcb+07}.
The \ion{N}{5} equivalent widths and line centroids are
not observed to vary with statistical significance.  We
set upper limits of $\Delta W < 25$m\AA\ and $\Delta v = 10 \mkms$
based on this modest S/N spectra.

\clearpage

\begin{deluxetable}{lcc}
\tablewidth{0pc}
\tablecaption{VOLUME DENSITY ESTIMATES\label{tab:phys}}
\tabletypesize{\footnotesize}
\tablehead{\colhead{GRB} &
\colhead{[N/H]$^a$} & \colhead{$\log (n_H/\cm{-3})^b$} }
\startdata
GRB021004&$-3.0^*$&$ 2.3$\\
GRB030323&$-1.9^*$&$< 0.9$\\
GRB050730&$-3.2$&$ 1.9$\\
GRB050820&$>-1.3$&$<-0.5$\\
GRB050922C&$<-4.1$&$> 3.0$\\
GRB060206&$-1.9^*$&$ 0.3$\\
GRB060607&$-2.0^*$&$<-0.7$\\
\enddata
\tablenotetext{a}{Nitrogen metallicity inferred from the ratio of N$^0$ and H$^0$ column densities \citep{pcd+07}.  This gas is located at a distance of 100pc to a few kpc from the GRB afterglow.  Systems marked with a * do not have \ion{N}{1} observations and we have set [N/H]=[M/H]-1.}
\tablenotetext{b}{Scaled from our photoionization models assuming
the afterglow (GRB050730) used throughout the paper.}
 
\end{deluxetable}

\section{Concluding Remarks}

We have performed a survey of \ion{N}{5} absorption along seven GRB
sightlines and reported six positive detections within 100\kms\ of the
neutral gas associated with the host galaxy.  Aside from the
GRB~050820 sightline (where the \ion{N}{5} absorption is broad, weak
and offset by $\delta v \approx -100\mkms$), the \nion\ gas has large
column density and kinematically `cold' line-profiles.  The latter
characteristic refers to a low velocity dispersion and a small offset
$|\delta v| < 20 \mkms$ from the neutral gas.  The \ion{N}{5} profiles
are also coincident in velocity 
with fine-structure absorption which suggests
the gas is located within $\approx 1$\,kpc of the GRB afterglow.

We have explored several scenarios that could produce \ion{N}{5}
absorption along GRB sightlines.  Models related to the halo of the
host galaxy or material shock-heated by the progenitor's stellar wind
are disfavored by the observations.  In contrast, a scenario where the
\nion\ gas is material photoionized by the GRB afterglow naturally
reproduces the observations provided the gas at $r \approx 10$\,pc is
cold ($T \approx 10^4$\,K), and has a modest density ($n \approx 1
\cm{-3}$), a non-negligible metallicity ([N/H]~$>-2$), and a similar
velocity as the ISM at $r \gtrsim 100$\,pc.

The afterglow photoionization model places several important
constraints on the progenitor and its environment.  In particular,
this scenario requires that the stellar wind of the progenitor
terminates at less than $r \approx 10$\,pc.  This suggests the GRB
progenitor has a weak, main-sequence stellar wind owing to a low
mass-loss rate, a low wind speed, and/or a short lifetime.  These
characteristics may be a natural consequence of the progenitors that
favor the GRB phenomenon, e.g.\ higher mass, lower metallicity stars.
The wind can also be confined by introducing a dense 
$(n \gtrsim 10^3 \cm{-3})$ external medium.  Perhaps GRBs are
preferentially embedded within the dense regions of molecular clouds
as opposed to a violent, starburst region.  On the other hand, we note
that the \ion{N}{5} absorption detected along the GRB~050820 sightline
has characteristics consistent with shock-heated gas provided a shock
with $v \approx 100\mkms$.  In this case, we may be seeing the
signatures of a starburst galaxy.

Before concluding, we wish to comment on a few directions for future
research.  One aspect to explore is how the soft X-ray absorption
observed in the afterglow spectroscopy compares with the
\nion\ observations.  To zeroth order, both measurements are sensitive
to the column density of metals near the GRB progenitor, although
likely at somewhat different radii.  A comprehensive model of these
observations may constrain the density profile of gas close to the
GRB.  Another implication of our research is that one predicts
recombinations of the \nion\ gas (and other high-ions, e.g.\ O$^{+5}$)
once the afterglow fades.  For low $z$ GRB, it may be possible to
observe this line emission with a sensitive ultraviolet telescope
\citep{prl00}.  Finally, one predicts that other high ionization
states will be produced by the afterglow (e.g.\ O$^{+5}$, S$^{+5}$)
that could be studied in a similar fashion to constrain the relative
abundances of gas near the progenitor star.

\acknowledgements We acknowledge helpful discussions with D. Kasen and
A.-J. van Marle.  We thank D. Whalen and A. Heger for their
calculations of the photon flux from massive stars.  We thank
S. Savaglio for providing her reduction of the VLT/UVES cb58 data for
analysis.  J. X. P. is partially supported by NASA/Swift grants
NNG06GJ07G and NNX07AE94G and an NSF CAREER grant (AST-0548180).
Based on observations made with ESO Telescopes at the Paranal
Observatories and accessed from the ESO data archive.

%\bibliographystyle{../../../Bibli/apj}
%\bibliography{../../../Bibli/allrefs}

%\clearpage

%\input{../Tables/tab_grbobs.tex}
%\input{../Tables/tab_summ.tex}
%\input{../Tables/tab_phys.tex}

\end{document}